\documentclass[preprint,5p,times,twocolumn]{elsarticle}
\usepackage{amssymb}
\usepackage{amsmath}
\usepackage{float}
\journal{Physics Letters B}

\begin{document}

\begin{frontmatter}



\title{Limits on new strongly interacting matter from measurements of Transverse Energy-Energy Correlations at $\sqrt{s} = 13$ TeV at the LHC}


\author[add1]{Javier Llorente}
\ead[add1]{javier.llorente.merino@cern.ch}
\author[add1]{Eva Sánchez}
\ead[add2]{eva.sanchez@ciemat.es}
\address[add1]{CIEMAT. Avenida Complutense 40, 28040. Madrid, Spain.}
%\address[add2]{Facultad de Ciencias Físicas, Universidad Complutense de Madrid. Madrid, Spain.}

%% Abstract
\begin{abstract}
This work establishes 95\% confidence level limits to models incorporating additional fermions sensitive to the strong interaction. Precision measurements of Transverse Energy-Energy Correlations at the ATLAS experiment are used, exploiting their dependence on the strong coupling constant to analyse the effects of introducing new fermions with colour charge on the Renormalisation Group Equation. The comparison between theoretical predictions, corrected up to next-to-next-to-leading order, and the data collected by ATLAS at $\sqrt{s} = 13$ TeV allows to constrain physics models proposing the existence of new fermions with masses up to 4 TeV, independently of assumptions on the fermion decay.
\end{abstract}



%% Keywords
%\begin{keyword}



%\end{keyword}

\end{frontmatter}



\section{Introduction}
\label{sec:intro}

After decades of experiments that have tested the Standard Model of particle physics (SM), it has been established as the model that most accurately describes both the elementary particles known to date and the forces that determine the interaction between these particles, with the exception of gravity. However, the SM is known to be incomplete for several reasons. Among them, the matter-antimatter asymmetry in the Universe, the lack of explanation for dark matter, and the non-inclusion of General Relativity in the SM are the main incentives for searching for physics beyond the Standard Model (BSM)~\cite{doi:10.1098/rsta.2015.0259}.\\
\newline
Today, some of the experiments that focus part of its efforts on the search for BSM physics~\cite{ATLAS:2024rcu,ATLAS:2024kgp,ATLAS:2024rzi,CMS:2021knz,CMS:2024gyw,CMS:2024gxp} are installed at the Large Hadron Collider (LHC), the most powerful particle collider to date, capable of reaching centre-of-mass energies of 13.6 TeV~\cite{Lyndon_Evans_2008}. Since the existence of new particles could leave signals at the energies available at the LHC, indications of new physics would be indirectly detectable at this collider by performing precise measurements of observables whose theoretical prediction is modified by reformulating the SM. Because the colliding objects at the LHC are protons, it is very common for their constituents to interact strongly to give rise to final states that contain jets, i.e., sets of high-energy collimated particles resulting from the fragmentation of quarks or gluons that group together forming colour-neutral hadrons~\cite{Cacciari_2012}. The interactions between particles with colour charge are theoretically described in the SM by a quantum field theory called Quantum Chromodynamics (QCD). Therefore, one way to check the validity of the SM can be through the examination of QCD as a perturbative theory by analyzing observables measured in events with multiple jets in the final state~\cite{eventShapes,2023}. Specifically, observables sensitive to how the strong coupling constant $\alpha_s$, which determines the strength of interactions between quarks and gluons, varies with the interaction scale, $Q^2$, and with the theoretical model, can be analysed to restrict BSM models that include new fermions with colour charge~\cite{Becciolini_2015,Llorente_2018}. \\
\newline
Transverse Energy-Energy Correlations (TEEC)~\cite{ALI1984447}, i.e. the weighted relative azimuth between all pairs of jets in a given event, constitute an observable of particular interest because of its experimental precision, since they can be measured to $\mathcal{O}(3\%)$~\cite{2023}. They also exhibit excellent perturbative convergence properties, since NNLO corrections to this observable are of $\mathcal{O}(1\%)$~\cite{Alvarez_2023}, as well as a strong dependence on $\alpha_s$~\cite{2023, Ali_2012}. Therefore, modifying the $\alpha_s(Q^2)$ function by introducing new fermions with colour charge in the SM implies a different theoretical prediction for the TEEC in BSM models with respect to the SM. Consequently, the analysis of TEEC allows to set limits on the existence of new fermions independently of any decay model, as it only requires a coupling model to the strong sector of the Standard Model. This study has been previously conducted~\cite{Llorente_2018} using theoretical predictions of TEEC at next-to-leading order (NLO) in perturbative QCD (pQCD) and data collected at the LHC for centre-of-mass energies of $\sqrt{s}=$ 8 TeV~\cite{Aaboud_2017}. The aim of the present work is to extend this analysis by considering theoretical predictions to the next perturbative order in QCD, which have been recently calculated in Ref.~\cite{Alvarez_2023}, and data taken at higher energies, $\sqrt{s} =$ 13 TeV~\cite{2023}, in order to improve the limits on BSM models including new fermions with colour charge, depending on their masses and the representation of SU(3) under which these fermions transform. Examples of such models include, for instance, supersymmetric models~\cite{WessZumino1974_1,WessZumino1974_2,FerraraZumino1974,Salam1974} including a gluino, $\tilde{g}$, or dark matter models with dark quarks coupling to the Standard Model gluons~\cite{cohen2017}.\\
\newline
The report is organised as follows. Section~\ref{alpha_s} introduces the theoretical background, including the  modification of the strong coupling constant when new fermions with colour charge are added to the SM. Section~\ref{teec} defines Transverse Energy-Energy Correlations and, subsequently, Section~\ref{dataset} details the dataset and theoretical predictions used. The analysis concludes with the determination of limits on BSM models in Section~\ref{pvalue}. Finally, the conclusions of the study are presented in Section~\ref{conclusions}.

\section{Quantum Chromodynamics and the Strong Coupling Constant}
\label{alpha_s}

\begin{figure}[!h]
  \centering
  \includegraphics[width=0.49\textwidth]{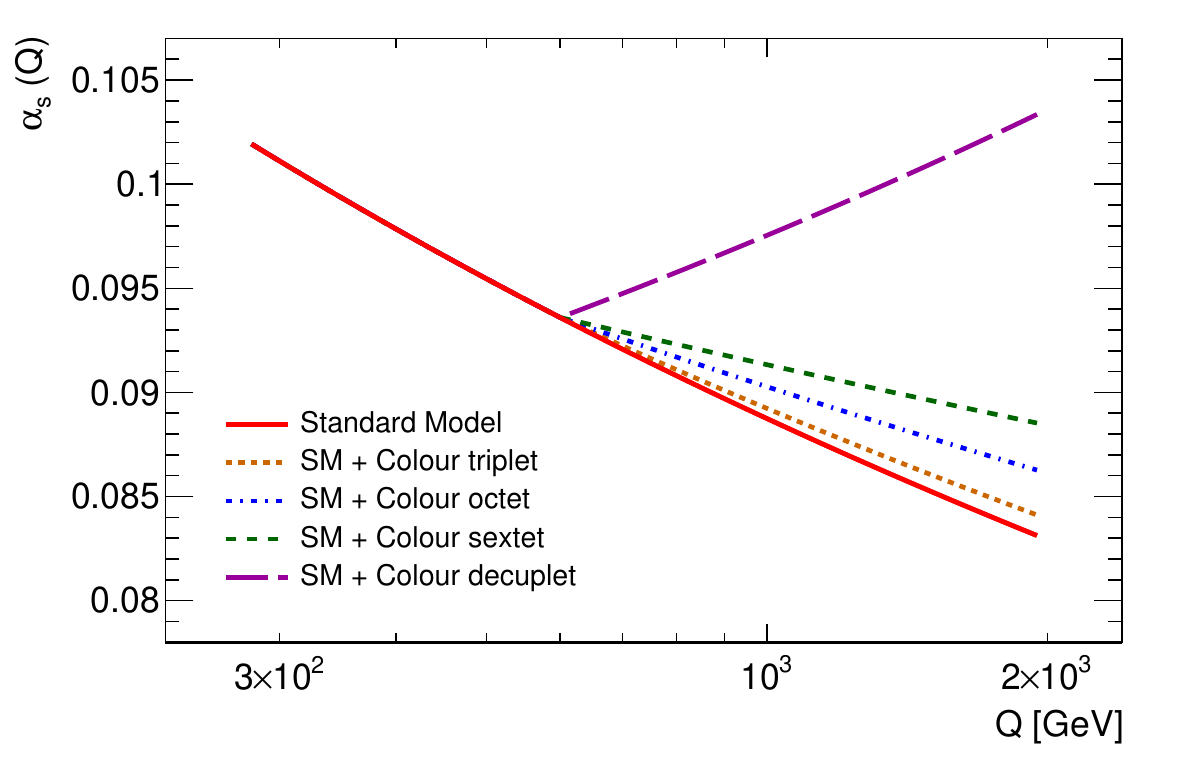}
  \caption{Variation of $\alpha_s$ with the renormalisation scale, calculated at NLO for different BSM models, introducing a new colour-charged fermion into the SM. The resulting curves are shown for a fermion with mass $m_X = 600$ GeV, transforming as a colour triplet, octet, sextet and decuplet, respectively.}
    \label{fig:alphaS}
\end{figure}

\begin{figure}[!h]
    \centering
    \includegraphics[width=0.49\textwidth]{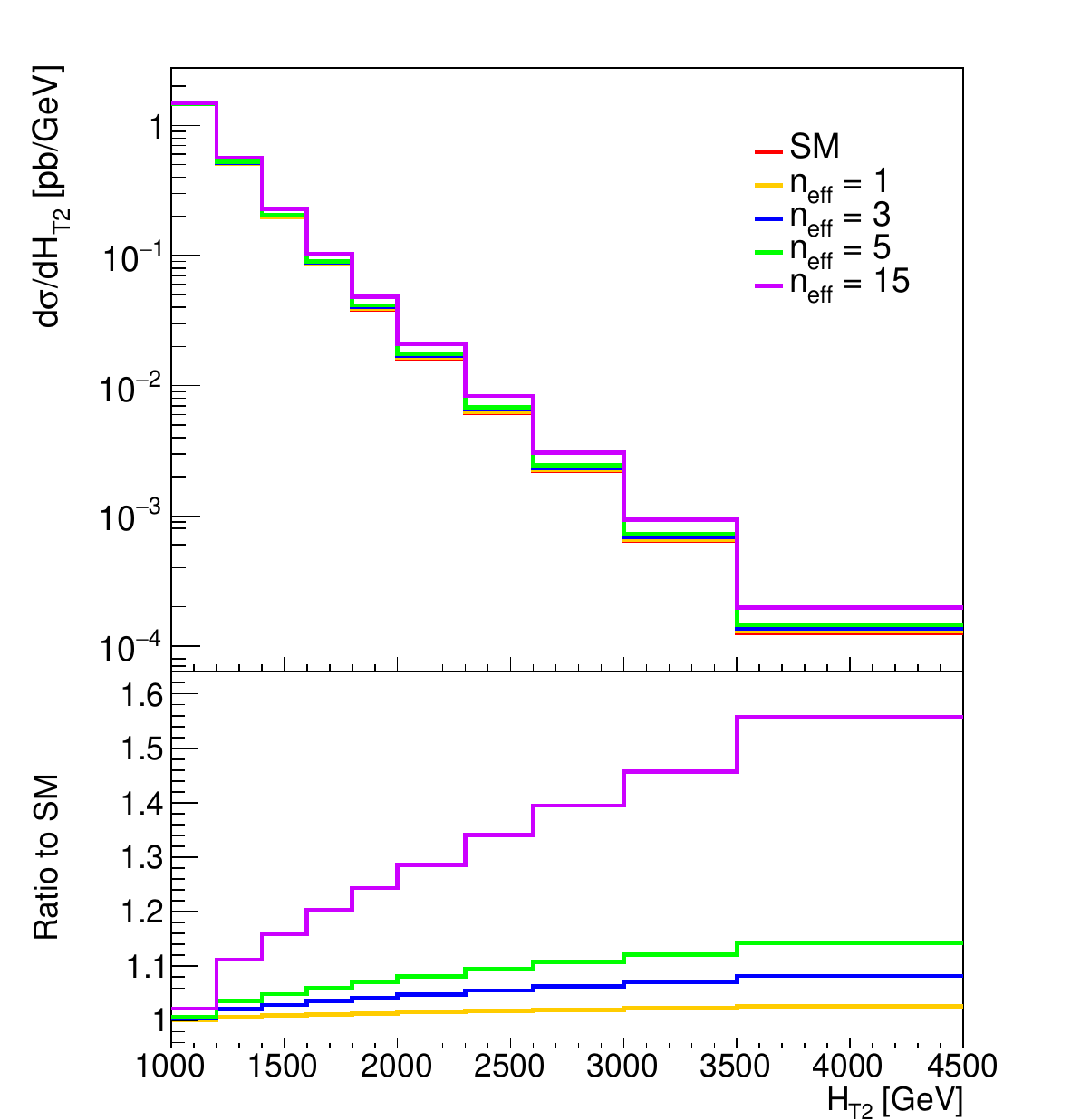}
    \caption{Differential cross section of the process $pp \rightarrow \mbox{jj}$ as a function of $H_{T2}$, calculated at NLO for BSM models introducing a new fermion with mass $m_X = 600$ GeV and $n_{\text{eff}} = 1, 3, 5$ and 15, respectively. On the lower panel, the ratio of the theoretical predictions for BSM models to the expected values for the SM is shown.}
    \label{fig:cross_section}
\end{figure}

Quantum Chromodynamics (QCD) is a non-abelian gauge theory whose Lagrangian is characterised by its invariance under the special unitary group SU(3). The renormalisation of the coupling constant of the theory, $\alpha_s$, requires the Renormalisation Group Equation (RGE) to hold:
\begin{equation}
  \frac{\partial \alpha_s}{\partial \log Q^2} = \beta(\alpha_s) = -\alpha_s^2(\beta_0 + \beta_1 \alpha_s + \mathcal{O}(\alpha_s^2)),
  \label{eq:rge}
\end{equation}
where
\begin{equation}
  \beta_0 = \frac{1}{4\pi} \left( 11 - \frac{2}{3} n_f \right); \ \ \ \ \ \beta_1 = \frac{1}{(4\pi)^2} \left( 102 - \frac{38}{3} n_f \right),
  \label{eq:beta_SM}
\end{equation}
and $n_f$ is the number of active flavours at the considered energy.\\
\newline
If new colour-charged fermions are included, it is necessary to modify the definition of the coefficients $\beta_0$ and $\beta_1$. In Ref.~\cite{Becciolini_2015}, these parameters are calculated for BSM models with new fermions sensitive to the strong interaction,
\begin{eqnarray}
  \beta_0 = \frac{1}{4\pi} \left( 11 - \frac{2}{3} n_f - \frac{4}{3} n_X T_X \right)\label{eq:beta0}\\
  \beta_1 = \frac{1}{(4\pi)^2} \left[ 102 - \frac{38}{3} n_f - 20 n_X T_X \left( 1 + \frac{C_X}{5} \right) \right]\label{eq:beta1}
\end{eqnarray}
where $n_X$ represents the number of new fermions; $T_X$ is the group factor given by $\text{Tr}(t^A t^B) = T_X\delta^{AB}$, where $t^A$ are the generators of SU(3) in a given representation; and $C_X$ is the Casimir of that representation, fulfilling $\sum_A t^A_{ab} t^A_{bc} = C_X\delta_{ac}$. For SU(3) triplets, i.e. quarks, $T_X=1/2$ and $C_X=4/3$. In what follows, the parameter $n_{\text{eff}} = 2 \sum n_X T_X$ and the mass of the new fermion, $m_X$, will be used to characterise a specific BSM model. \\
\newline
The solution to the Renormalisation Group Equation provides the way in which the strong coupling constant varies with the renormalisation scale. At NLO, it reads:
\begin{equation}
  \alpha_s(Q^2) = \frac{1}{\beta_0 \log z} \left[ 1 - \frac{\beta_1}{\beta_0^2} \frac{\log(\log z)}{\log z} \right], ~~~ z = \frac{Q^2}{\Lambda^2},
  \label{eq:alpha_s}
\end{equation}
where $\Lambda$ is the integration constant defined as
\begin{equation}
    \log \frac{Q^2}{\Lambda^2} = -\int_{\alpha_s(Q^2)}^{\infty} \frac{\text{d}x}{\beta(x)}.
\end{equation}
\newline
The definition of coefficients $\beta_0$ and $\beta_1$ introduces a dependence of the strong coupling constant evaluated at a specific energy on the specific BSM model. Figure~\ref{fig:alphaS} shows the running of $\alpha_s$ with the renormalisation scale for different BSM models introducing a new colour-charged fermion. The curves shown are calculated at NLO in the strong coupling constant and, for illustrative purposes, the mass of the new fermion is fixed at 600 GeV. The represented models correspond to $n_{\text{eff}} = 1$ (SM + colour triplet), $n_{\text{eff}} = 3$ (SM + colour octet), $n_{\text{eff}} = 5$ (SM + colour sextet), and $n_{\text{eff}} = 15$ (SM + colour decuplet). As observed in the figure, the $\alpha_s$ curve as a function of $Q$ begins to deviate from the SM curve at values of $Q$ equal to the mass of the new fermion. For higher $n_{\text{eff}}$, the separation of the curve with respect to the SM becomes more pronounced, and for some models, even asymptotic freedom ($\alpha_s \rightarrow 0$ as $Q \rightarrow \infty$) is lost. This phenomenon occurs at first order when the derivative of $\alpha_s(Q^2)$ vanishes. For models with 6 active flavours, this happens for $n_{\text{eff}} \geq 10.5$. Thus, in the graph, the only represented model that loses asymptotic freedom is that with an additional fermion transforming as a colour decuplet, i.e., $n_{\text{eff}} = 15$. It should be noted that, given a fermion of mass $m_X$, the deviation of $\alpha_s$ with respect to the SM curve is greater for higher energy scales.
\noindent
The modification of $\alpha_s(Q^2)$ caused by the introduction of new fermions is reflected on measurable quantities at the LHC by means of their perturbative expansion. One observable that allows to understand how the theoretical predictions are changed by introducing such fermions is the two-jet production cross section in $pp$ collisions, $\sigma(pp \rightarrow \mbox{jj})$, as it is calculated perturbatively as a power series in $\alpha_s$~\cite{dijet2018}. Figure~\ref{fig:cross_section} shows the differential cross section of the process $pp \rightarrow \mbox{jj}$ as a function of the scalar sum of the transverse momenta of the two highest-$p_T$ jets, $H_{T2} = p_{T1} + p_{T2}$. The SM predictions, as well as those introducing a new strongly interacting fermion with a mass of 600 GeV and $n_{\text{eff}} = 1,3,5$ and 15 into the SM, are included for illustrative purposes. The SM cross section is represented in red. On the lower panel, the ratios of the predictions for each BSM model to the corresponding SM prediction are shown. It can be seen that the existence of massive new fermions deviates the theoretical prediction for the cross section from the SM prediction at the energies available at the LHC. Although the trend of the curves is the same for the represented BSM models (the cross section decreases as the energy of the process increases) the cross sections for models with higher $n_{\text{eff}}$ take larger values, and the difference with the SM becomes greater at higher energies.

\section{Transverse Energy-Energy Correlations}
\label{teec}
Transverse Energy-Energy Correlations~\cite{ALI1984447} offer several advantages: they can be calculated in pQCD up to NNLO accuracy~\cite{Alvarez_2023} and, as introduced in Section~\ref{sec:intro}, they show excellent perturbative convergence properties as well as stability with respect to different experimental and theoretical effects. Moreover, this observable shows a strong dependence on the value of $\alpha_s$, which has allowed the ATLAS Collaboration to determine $\alpha_s(m_Z)$ from them at different centre-of-mass energies ~\cite{Aad_2015, Aaboud_2018, 2023}. The mathematical expression defining TEEC is~\cite{Altarelli1994}:
\begin{equation}
  \frac{1}{\sigma}\frac{\text{d}\Sigma}{\text{d}\cos\phi} \equiv \frac{1}{\sigma}\sum_{ij}\int\frac{\text{d}\sigma}{\text{d}x_{T_i}\text{d}x_{T_j}\text{d}\cos\phi}x_{T_i}x_{T_j}\text{d}x_{T_i}\text{d}x_{T_j},
  \label{eq:teec}
\end{equation}
where $\sigma$ is the cross-section for two-jet production and $x_{T_i}$ is the fraction of energy carried by jet $i$ relative to the total,
\begin{equation}
    x_{T_i} = \frac{E_{T_i}}{\sum_k E_{T_k}}.
\end{equation}
\noindent
At NLO in perturbative QCD, TEEC can be expressed as a second-order expansion in powers of $\alpha_s(Q^2)$ as~\cite{Ali_2012}
\begin{equation}
  \frac{1}{\sigma}\frac{\text{d}\Sigma}{\text{d}\cos\phi}\propto \frac{\alpha_s(Q^2)}{\pi}F(\phi)\left[1+\frac{\alpha_s(Q^2)}{\pi}G(\phi)\right],
  \label{eq:pTEEC}
\end{equation}
where $F(\phi)$ and $G(\phi)$ are functions to be determined. The TEEC distributions exhibit two peaking structures at $\cos\phi\simeq -1$ and $\cos\phi\simeq +1$. The former corresponds to back-to-back dijet configurations at $\Delta\phi \simeq \pi$, while the latter corresponds to self-correlations and collinear emissions at $\Delta\phi \simeq 0$. Intermediate values of $\cos\phi$ correspond to radiation emitted at larger angles. Thus, the first non-trivial contribution to the TEEC cross section corresponds to leading-order three-jet production.

\section{Dataset and Theoretical Predictions}
\label{dataset}
The dataset used for the analysis was collected by the ATLAS experiment during the LHC Run II, between 2015 and 2018. The data were taken at a centre-of-mass energy of $\sqrt{s} =$ 13 TeV, with a total integrated luminosity of 139 fb$^{-1}$, and corresponds to the distributions published in Ref.~\cite{2023}. \\
\newline
Theoretical predictions were obtained using the \textsc{NLOJet++} program~\cite{Nagy_2003}, which allows for the calculation of the NLO cross sections for three-jet production, $pp \rightarrow \mbox{jjj}$. In total, distributions for the TEEC are calculated for 400 BSM models, modifying the QCD $\beta$-function as in Eq.~\ref{eq:beta0} and~\ref{eq:beta1} by varying $n_{\text{eff}}$ from 0 to 20 and the mass $m_X$ up to 4 TeV. The values of the Casimir $C_X$ are set to 3 for all models. This choice is driven by the studies in Ref.~\cite{Becciolini_2015}, where it was shown that variations of $C_X$ between $4/3$ and $6$ produced a negligible effect on the exclusion limits. The parton distribution functions (PDF) used for our central result are those from MMHT2014~\cite{Harland_Lang_2015}, while the result is cross-checked using the CT18 PDF~\cite{Hou_2021} to give a quantitative idea of the dependence of the exclusion limits with the PDF choice. The central value of $\alpha_s=0.118$ is used in both PDF sets.
\begin{figure*}[!h]
  \begin{center}
    \includegraphics[width = 0.46\textwidth]{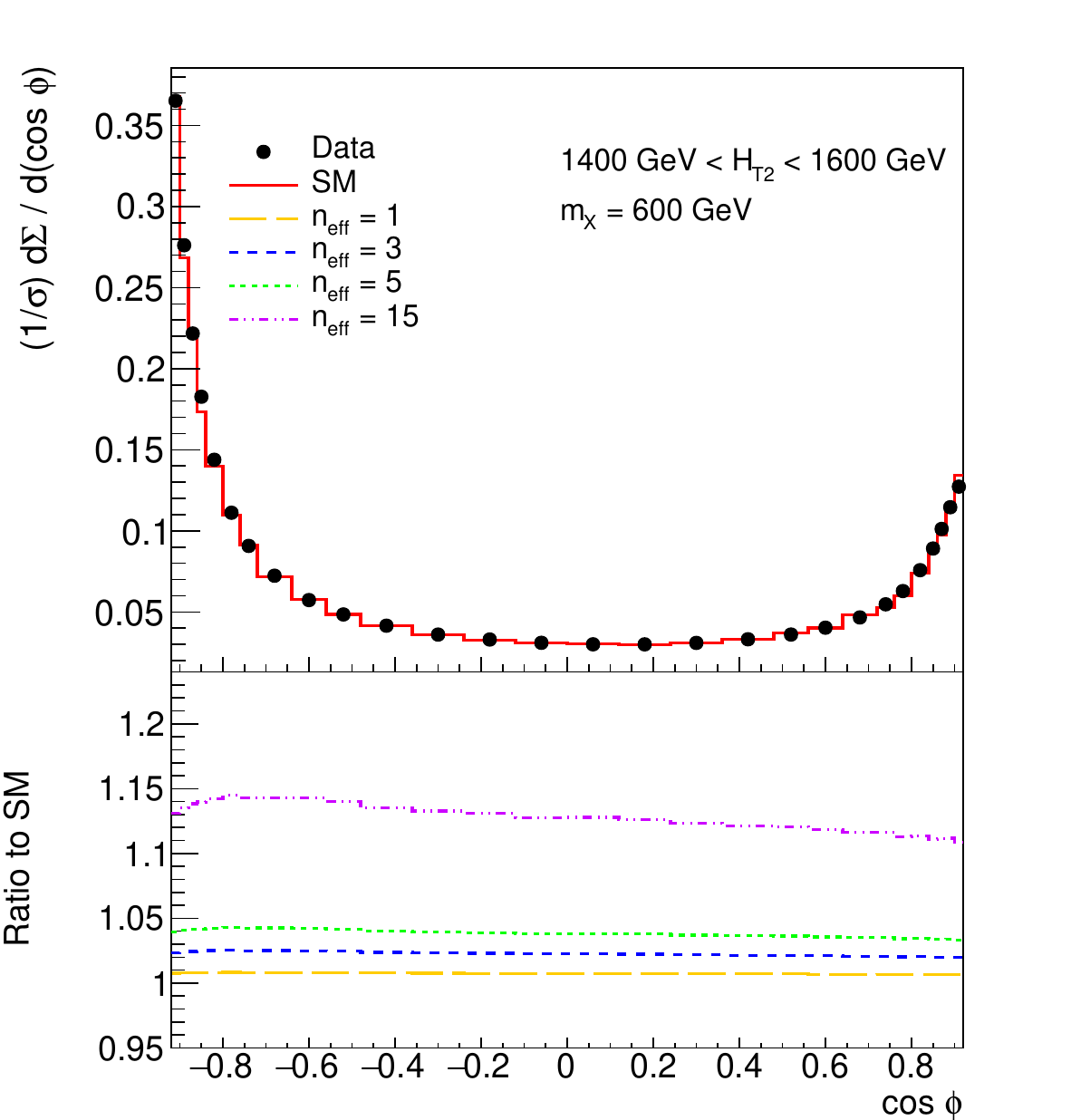}
    \includegraphics[width = 0.46\textwidth]{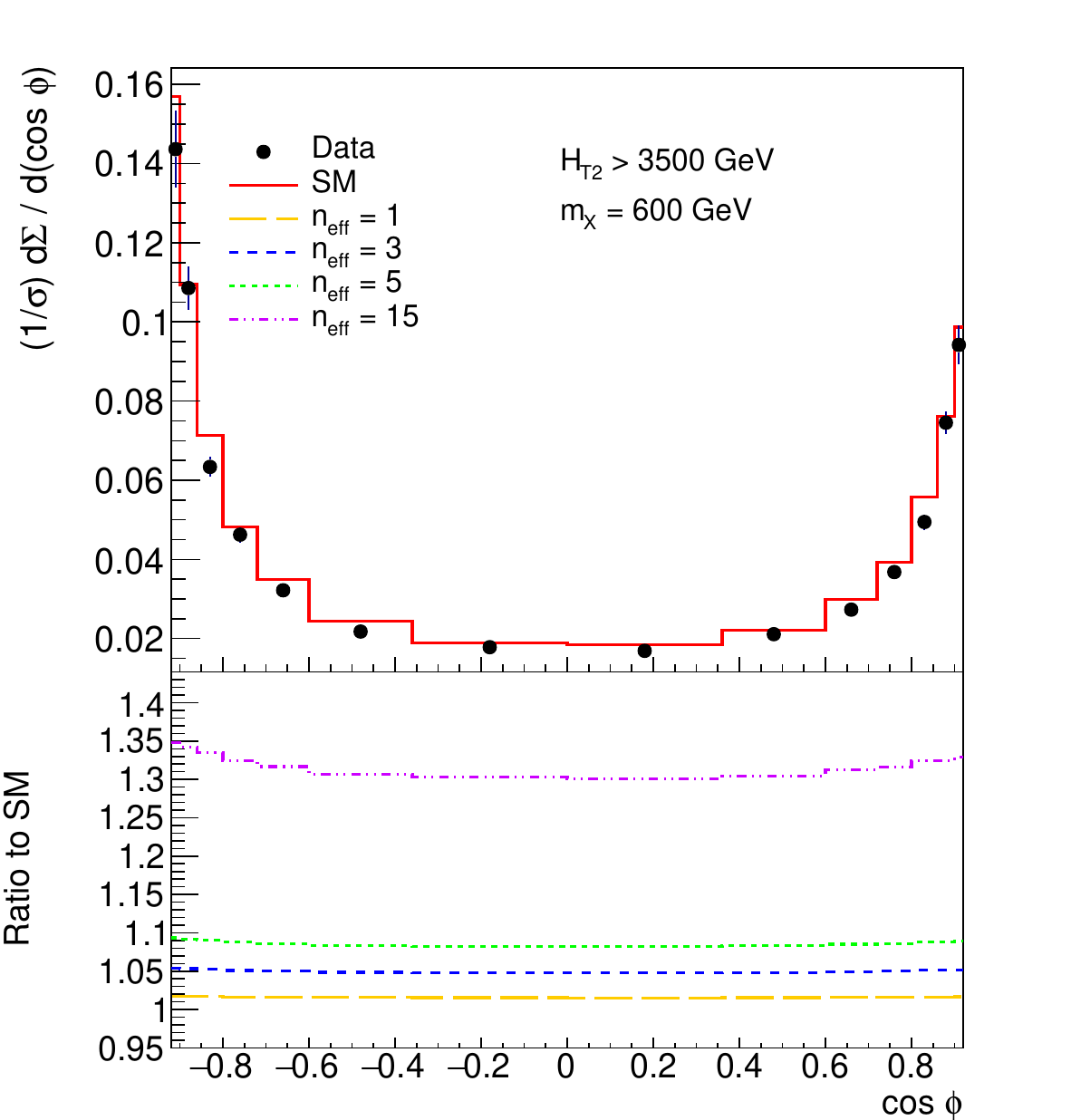}
  \end{center}
  \caption{Theoretical predictions of transverse energy-energy correlations for two $H_{T2}$ energy ranges. In each individual figure, the upper graph shows the approximate NNLO theoretical prediction of the SM and the data collected at 13 TeV. The lower graph represents the ratio of the prediction of four BSM models including a new fermion with colour charge, mass $m_X = $ 600 GeV, and $n_{\text{eff}} =$ 1, 3, 5, 15 to the prediction of the SM.}
  \label{fig:dataTheory}
\end{figure*}
The tools mentioned above allow for obtaining theoretical predictions at NLO for the TEEC. However, this report aims to work with predictions at a higher precision order, specifically at NNLO. Therefore, the following approximation has been made. Since SM predictions at NNLO have been calculated using both MMHT2014 and CT18 ~\cite{Alvarez_2023}, it is possible to estimate the NNLO prediction for BSM models approximately. To do this, the so-called $K$-factors, defined as the ratios of the NNLO prediction to the NLO prediction in the SM are computed. These depend both on $H_{T2}$ and the azimuthal angle between jet pairs. The BSM theoretical prediction at NNLO is obtained by multiplying the corresponding $K$-factor by the NLO prediction, thus providing an approximation for the NNLO result, i.e., for each $H_{T2}$ bin
\begin{equation}
\left(\frac{1}{\sigma}\frac{\text{d}\Sigma}{\text{d}\cos\phi}\right)_{\text{BSM}}^{\text{NNLO}} \simeq K_{\text{SM}}(\cos\phi, H_{T2}) \times \left(\frac{1}{\sigma}\frac{\text{d}\Sigma}{\text{d}\cos\phi}\right)_{\text{BSM}}^{\text{NLO}}.
\label{eq:kfactor}
\end{equation}
The event selection follows that in Ref.~\cite{2023}. The study is, thus, limited to final states with two or more jets with $p_T >$ 60 GeV, and $|\eta| < 2.4$. Additionally, only events fulfilling the condition $H_{T2} > 1000$ GeV are analysed. The TEEC distributions are studied in the phase space $|\cos\phi| <$ 0.92 to avoid non-perturbative effects due to hadronisation, as well as logarithmic corrections due to collinear gluon resummation~\cite{Banfi:2010xy}. Finally, the renormalisation and factorisation scales are chosen as $\mu_R = \mu_F = \hat{H}_T/2$, where $\hat{H}_T = \sum_i p_{T,i}$ is the scalar sum of $p_T$ of all partons in the final state. This choice represents a compromise between recent theoretical developments proposing a scale proportional to $\hat{H}_T$~\cite{Alvarez_2023, Czakon_2021}, where all final-state partons are involved, and the tendency in experimental measurements to use scales of the order of the average transverse momentum of the $2 \rightarrow 2$ process~\cite{Chatrchyan_2013, Aaboud_2018}. This allows for a fair comparison with the results obtained in Ref.~\cite{Llorente_2018}, where a scale of the same numerical order, $\mu_R = \mu_F = (p_{T1} + p_{T2})/2$, with $p_{T1}$ and $p_{T2}$ being the transverse momenta of the two highest momentum jets, was chosen.\\
\newline
%\begin{figure*}[!h]
%  \begin{center}
%    \includegraphics[width = 0.46\textwidth]{./bin3_TFM.pdf}
%    \includegraphics[width = 0.46\textwidth]{./bin10_TFM.pdf}
%  \end{center}
%  \caption{Theoretical predictions of transverse energy-energy correlations for two $H_{T2}$ energy ranges. In each individual figure, the upper graph shows the approximate NNLO theoretical prediction of the SM and the data collected at 13 TeV. The lower graph represents the ratio of the prediction of four BSM models including a new fermion with colour charge, mass $m_X = $ 600 GeV, and $n_{\text{eff}} =$ 1, 3, 5, 15 to the prediction of the SM.}
%  \label{fig:dataTheory}
%\end{figure*}
Figure~\ref{fig:dataTheory} shows the approximate NNLO theoretical predictions for the TEEC as a function of the azimuthal angle between jet pairs, for two $H_{T2}$ ranges and four sample BSM models. Specifically, the displayed predictions were calculated for new fermions with a mass of 600 GeV and $n_{\text{eff}} = 1,3,5$ and 15. In the upper panel of each figure, the theoretical SM prediction (at exact NNLO) and the data from Ref.~\cite{2023} are shown. In the lower panels, the ratio of the BSM model prediction to the SM prediction is represented. As can be seen in the figures, the agreement of the Standard Model with the data collected at the LHC is, in general, very good.\\
\newline
On each of the subfigures, it can be observed that the ratio of the BSM models to the SM approaches unity as the value of $n_{\text{eff}}$ decreases. This is theoretically expected, as the coefficients $\beta_0$ and $\beta_1$ for BSM models (see Eq.~\ref{eq:beta0} and~\ref{eq:beta1}), which play a crucial role in defining $\alpha_s(Q^2)$ at NLO, differ from the coefficients for the SM (Eq.~\ref{eq:beta_SM}) on a term which is linear in $n_{\text{eff}}$. Thus, for low values of $n_{\text{eff}}$, the agreement with the SM is better. Another factor to highlight is that as the available energy in a process increases, BSM models diverge more from the Standard Model. This is because the introduction of a new fermion with mass $m_X$ modifies the function $\alpha_s(Q^2)$ starting from an energy equal to the mass of the new fermion onwards and, from that point, the $\alpha_s(Q^2)$ curves for different models begin to separate from the SM curve, with greater separation at higher energies (see Figure~\ref{fig:alphaS}).
\section{Limits on BSM models with new coloured fermions}
\label{pvalue}
\begin{figure}[!h]
  \centering
  \includegraphics[width=0.49\textwidth]{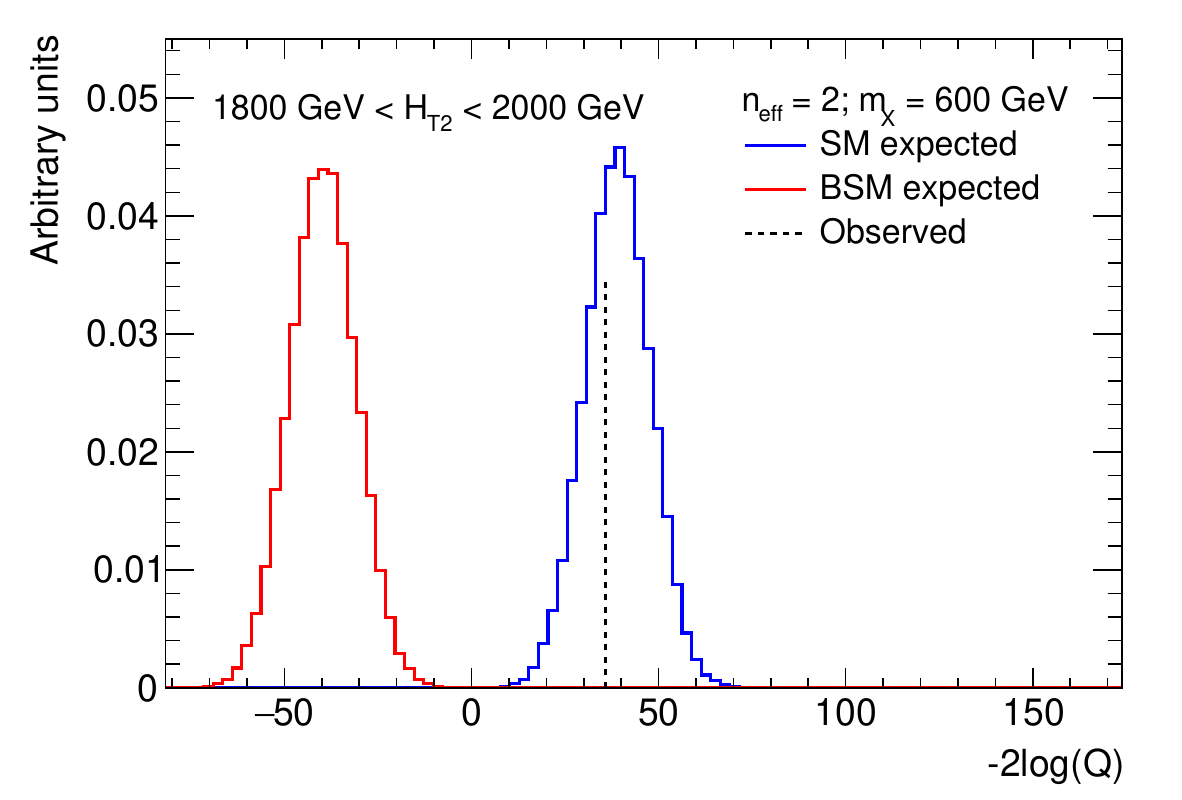}
  \caption{Distributions of the likelihood ratio $t=-2\log(Q)$, evaluated for the TEEC distribution with  1.8 TeV $< H_{T2} <$ 2 TeV, for the SM (blue) and BSM (red) hypotheses for a particular model with $n_{\text{eff}} = 2$ and $m_X = 600$ GeV. The observed value, $t_{\text{obs}}$, is also shown as a vertical dashed line.}
  \label{fig:tValues}
\end{figure}

\begin{figure*}[!h]
    \centering
    \includegraphics[width = 0.49\textwidth]{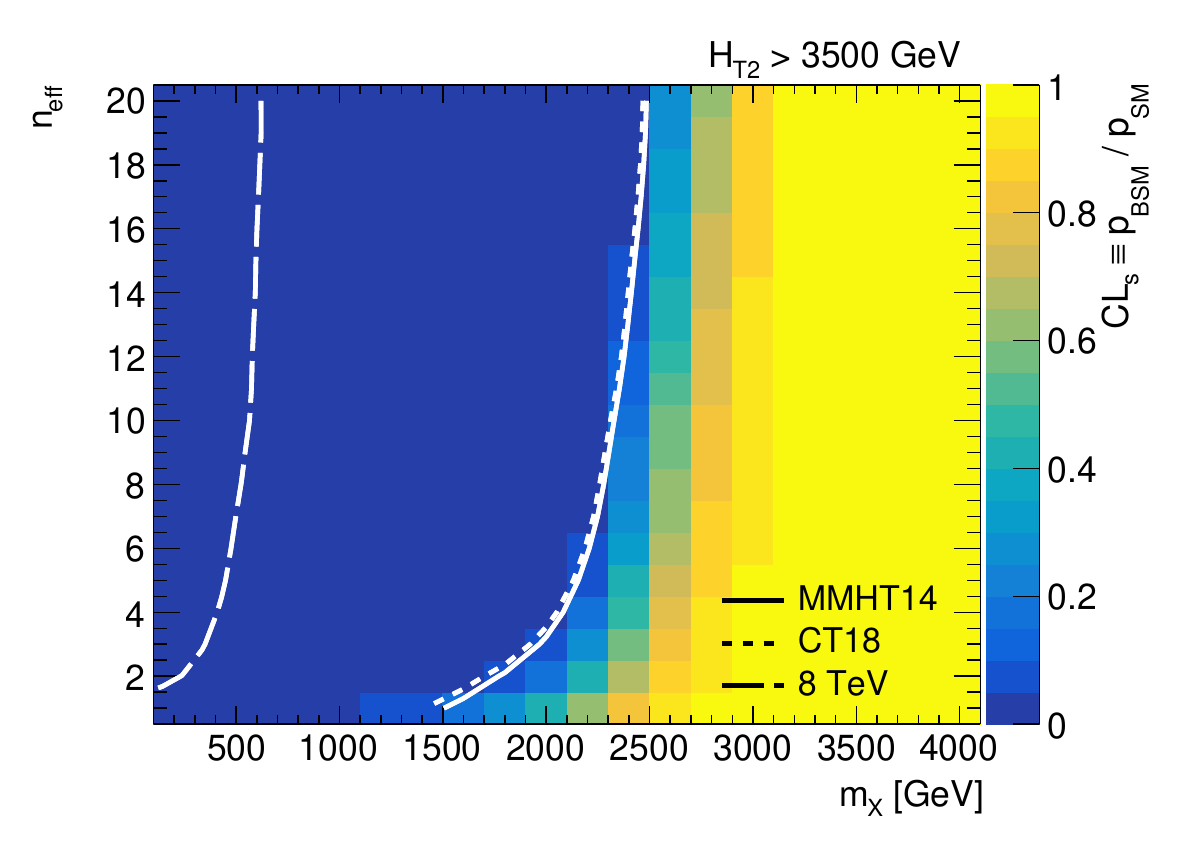}
    \includegraphics[width = 0.49\textwidth]{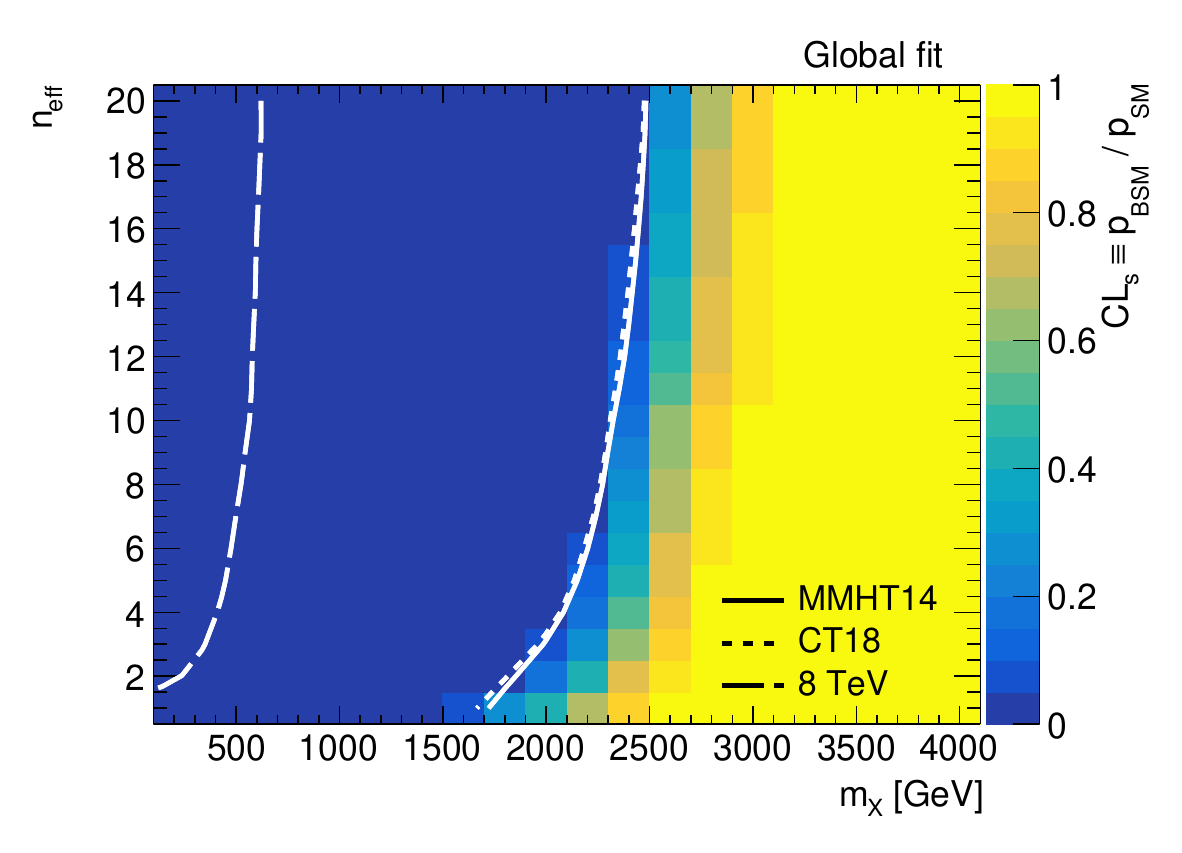}
    \caption{$\text{CL}_\text{s}$ for BSM models ($n_{\text{eff}}$, $m_X$) calculated from approximate NNLO predictions to TEEC and experimental data collected at 13 TeV. The left figure shows the $\text{CL}_\text{s}$ values obtained from a fit to the highest energy scale bin only, $H_{T2} >$ 3500 GeV, while the $\text{CL}_\text{s}$ values on right figure are obtained from a global fit to all $H_{T2}$ bins. To estimate the effect of the PDF on our analysis, the exclusion contours are obtained both for MMHT2014 and CT18, which show very similar results.}
    \label{fig:limitsFinal}
\end{figure*}

The comparison between the theoretical predictions and the experimental data allows us to exclude theoretical models whose predictions do not agree with experimental results. In order to quantify the agreement between data and theory, a likelihood ratio approach is used to find the $p$-value of a specific model with respect to the data, which allows us to understand the probability of such model to describe the experimental observations. The likelihood ratio is defined as
\begin{equation}
  t(x) = -2\log(Q) = -2 \log \frac{\mathcal{L}(x \ |\ H_{\text{BSM}})}{\mathcal{L}(x \ |\ H_{\text{SM}})}
  \label{eq:likRatio}
\end{equation}
where $\mathcal{L}(x \ |\ H_{\text{SM}})$ corresponds to the likelihood of the data distribution $x$ under the SM hypothesis (\emph{i.e.} background-only), while $\mathcal{L}(x \ |\ H_{\text{BSM}})$ is the likelihood of $x$ under the BSM hypothesis (\emph{i.e.} signal plus background). The expected distributions $f_i(t)$ under the SM and BSM hypotheses are estimated using pseudo-experiments which, given an observation $t_{\text{obs}} = t(x_\text{obs})$, are used to determine the $p$-values for each hypothesis as
\begin{eqnarray}
  p_\text{BSM} = \mathcal{P}(t\geq t_{\text{obs}} \ |\ H_{\text{BSM}}) = \int_{t_{\text{obs}}}^\infty f_{\text{BSM}}(t) dt\\
  p_\text{SM} = \mathcal{P}(t\geq t_{\text{obs}} \ |\ H_{\text{SM}}) = \int_{t_{\text{obs}}}^\infty f_{\text{SM}}(t) dt
\end{eqnarray}
The \text{CL}$_\text{s}$ test statistic is defined as the ratio of both $p$-values:
\begin{equation}
  \text{CL}_\text{s} = \frac{p_\text{BSM}}{p_\text{SM}}.
\end{equation}
A given BSM model is excluded if $\text{CL}_\text{s} < 0.05$.\\
\newline
The likelihood ratio in Eq.~\ref{eq:likRatio} is constructed using the $\chi^2$ definition used in the ATLAS measurement of the TEECs~\cite{2023}, which explicitly takes into account the correlations between the sources of the experimental uncertainties on the measured distributions,
\begin{equation}
    \chi^2(x\ |\ \alpha_s,\vec{\lambda}) = \sum_i\frac{(x_i - F_i(\alpha_s,\vec{\lambda}))^2}{\Delta x_i^2 + \Delta\tau_i^2} + \sum_k\lambda_k^2,
\end{equation}
where ${x_i}$ are the measured data for the TEEC, $\Delta x_i$ is the statistical uncertainty on these data points and $\Delta\tau_i$ is the statistical uncertainty on the theoretical predictions. The function $F_i(\alpha_s, \vec{\lambda})$ is defined as
\begin{equation}
    F_i(\alpha_s,\vec{\lambda}) = \psi_i(\alpha_s)\left(1+\sum_k\lambda_k\sigma^{(i)}_k\right),
\end{equation}
where the index $k$ runs over correlated sources of systematic uncertainty. In the above expression, $\psi_i(\alpha_s)$ is the analytical function parameterizing the dependence of TEEC on the strong coupling constant on bin $i$, $\{\lambda_k\}$ are nuisance parameters, and $\sigma^{(i)}_k$ is the $k$-th source of experimental uncertainty for bin $i$ in the measurement phase space. Under these conditions (and for fixed values of $\alpha_s$ and $\vec{\lambda}$), Eq.~\ref{eq:likRatio} reads
\begin{equation}
  t(x) = -2\log(Q) = \chi^2_\text{BSM}(x) - \chi^2_\text{SM}(x).
\end{equation}
Figure~\ref{fig:tValues} shows the distributions of $t=-2\log(Q)$ under the SM and BSM hypotheses for the particular model with $n_\text{eff} = 2$ and $m_X = 600$ GeV, as well as the observed value $t(x_\text{obs})$. The right-tail areas of both distributions taken from this value are used to define $\text{CL}_\text{s}$ which, for this particular case, has a value of 0 and the model is, therefore, excluded.\\
\begin{figure*}[!h]
  \centering
  \includegraphics[width = 0.49\textwidth]{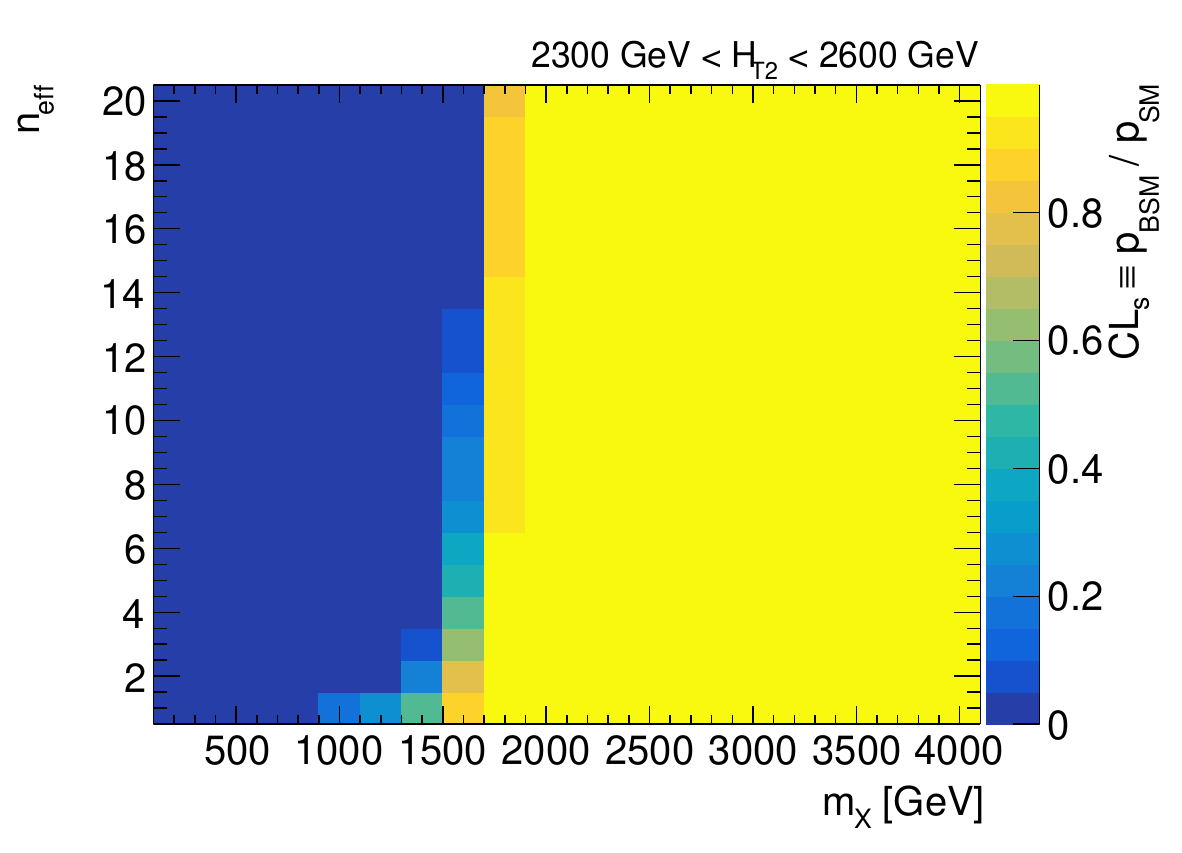}
  \includegraphics[width = 0.49 \textwidth]{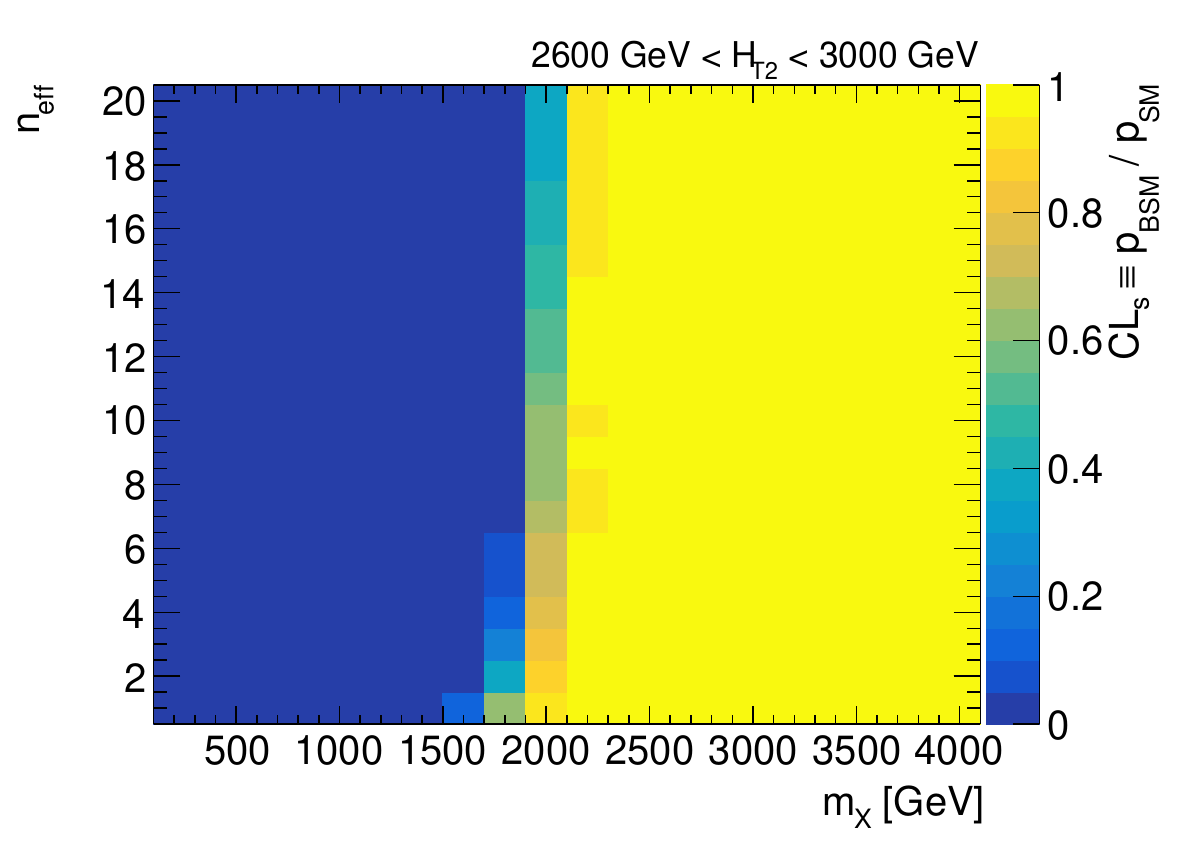}
  
  \includegraphics[width = 0.49 \textwidth]{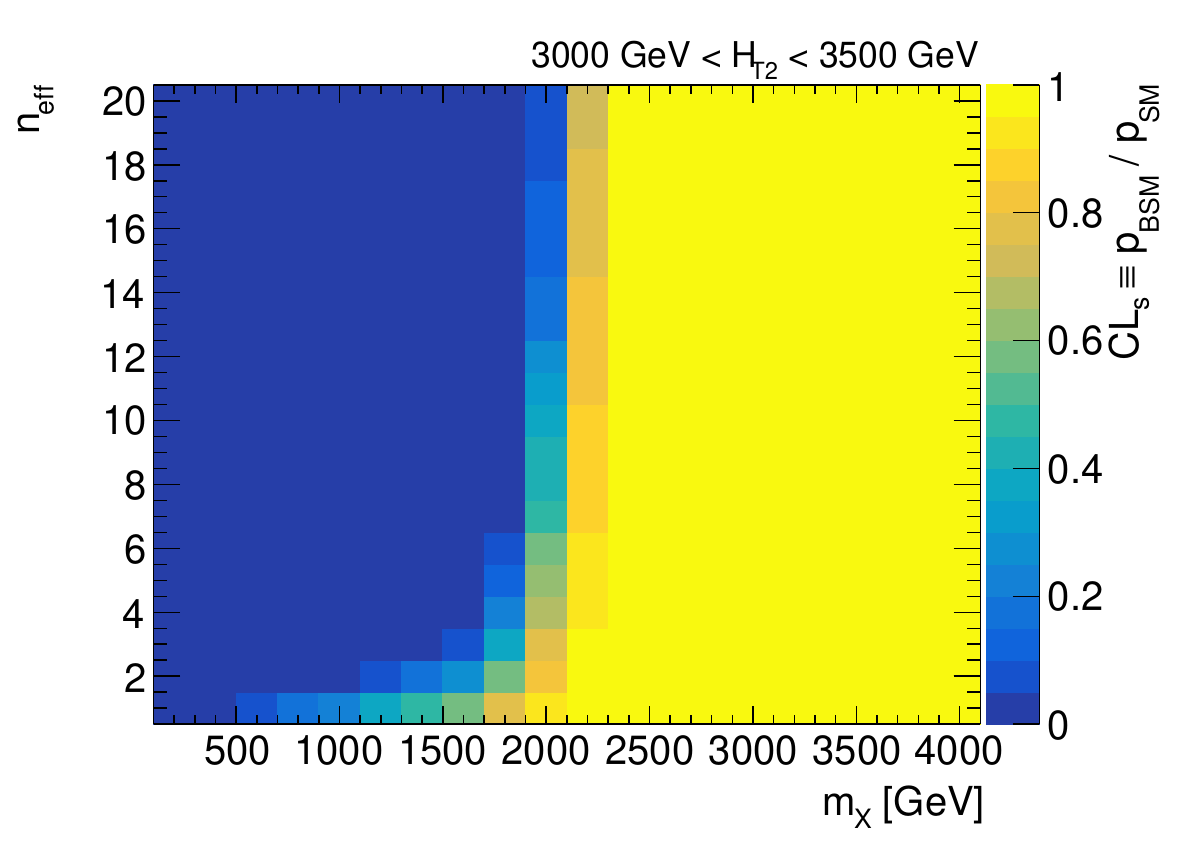}
  \includegraphics[width = 0.49 \textwidth]{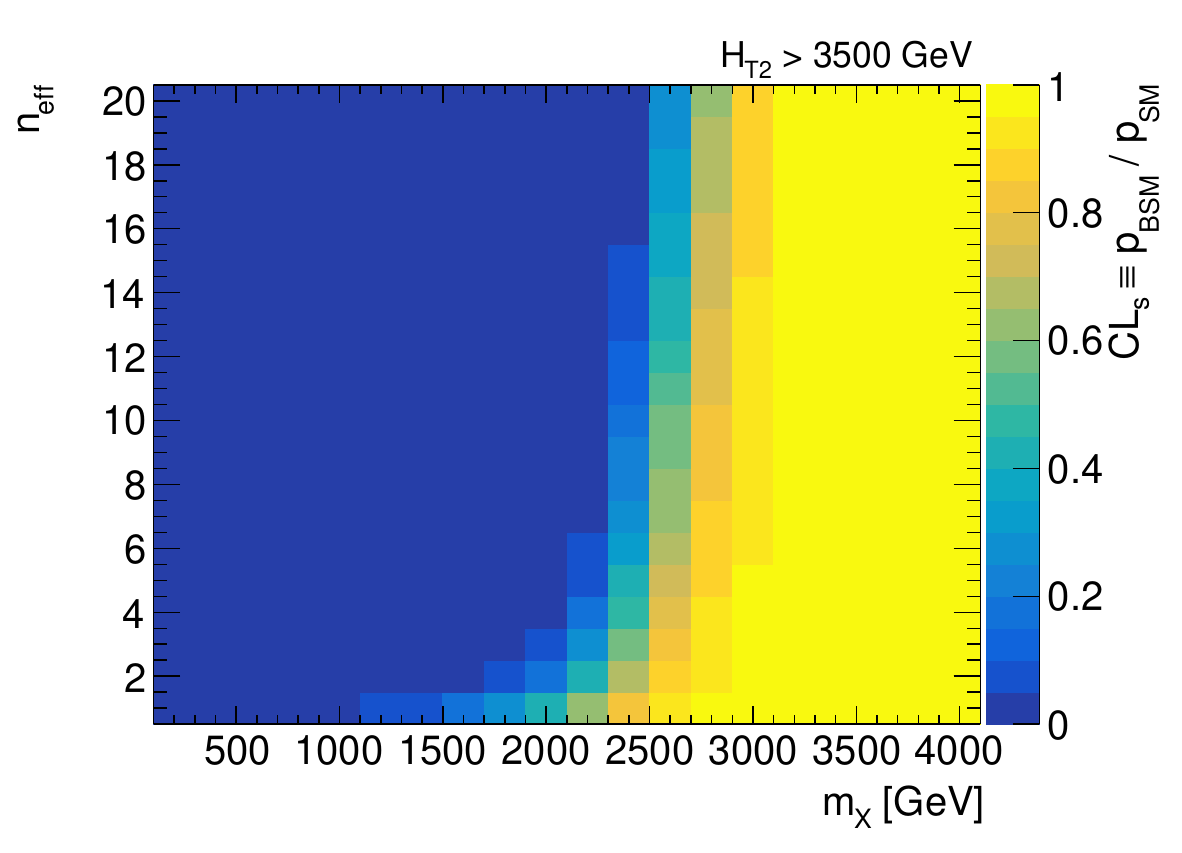}
  \caption{\text{CL}$_\text{s}$ for BSM models with $1 < n_\text{eff} < 20$ and 200 GeV $ < m_X <$ 4 TeV in 200 GeV intervals calculated with TEEC. The ($m_X$, $n_{\text{eff}}$) plane is shown for four energy ranges, 2.3 TeV $<$ $H_{T2}$ $<$ 2.6 TeV (top left), 2.6 TeV $< H_{T2} <$ 3.0 TeV (top right), 3.0 TeV $< H_{T2} <$ 3.5 TeV (bottom left) and $H_{T2} >$ 3.5 TeV (bottom right).}
  \label{fig:limits1}
\end{figure*}
\newline
\noindent
The value of $\text{CL}_\text{s}$ is estimated for 400 different BSM models with integer values of $n_{\text{eff}}$ ranging from 1 to 20 and fermion masses $m_X$ between 200 GeV and 4 TeV, in steps of 200 GeV. Figure~\ref{fig:limits1} shows the obtained results. In general, as can be seen in the figure, specific BSM models exhibit better agreement with experimental results for lower $n_{\text{eff}}$ and higher $m_X$. A comparison of the different figures shows that, as the energy scale is increased, fermions with larger masses can be excluded. This is not surprising, since, as mentioned earlier, the $\alpha_s(Q^2)$ curves for these models tend to diverge from the SM as $Q$ increases, and this effect is manifested in TEEC. Consequently, at higher energies, it is possible to exclude a greater number of BSM models.\\
\newline
Based on the results presented above, a 95\% confidence level contour can be obtained, separating the excluded region from the models that cannot be ruled out from the comparison of data to theory. To find such curve on the ($m_X$, $n_{\text{eff}}$) plane, the function that satisfies $\text{CL}_{\text{s}}$ = 0.05 is interpolated. As it is customary, this interpolation is done in the significance phase space, defined as $Z = \sqrt{2} \cdot \textup{erf}^{-1}(1 - 2\text{CL}_{\text{s}})$, where erf$^{-1}$ is the inverse error function. Figure~\ref{fig:limitsFinal} shows the corresponding $\text{CL}_\text{s}$ = 0.05 curve calculated with TEEC approximated at NNLO and experimental data collected at 13 TeV. On the left figure, the curve is obtained from the fit to the highest energy bin, $H_{T2} >$ 3500 GeV, while, on the right figure, a global fit is performed by considering all $(H_{T2},\cos\phi)$ bins together. The global fit is observed to have a similar exclusion power to the highest energy bin alone, although slightly improved for the lowest values of $n_\text{eff}$. While the solid line in Fig.~\ref{fig:limitsFinal} shows the curve for 13 TeV, the dashed line shows the curve for the 8 TeV results~\cite{Llorente_2018}. A clear improvement with respect to the 8 TeV results can be observed. In order to illustrate the smallness of the PDF dependence of the 95\% CL limits, the results are obtained for both the MMHT2014~\cite{Harland_Lang_2015} and the CT18~\cite{Hou_2021} PDF determinations. Figure~\ref{fig:limitsFinal} shows very similar results for both PDF choices. As discussed in some detail in Ref.~\cite{Llorente_2018}, the effect of varying the renormalisation and factorisation scales by a factor of 2 is not taken into account as a theoretical uncertainty. Since $\mu_R^2$ is taken as the argument of $\alpha_s(Q^2)$ in the perturbative calculation and BSM models with a massive fermion $X$ diverge from the SM curve only for $Q > m_X$, variations of $\mu_R$ by a factor of 2 are expected to cause a large variation on the 95\% CL limits on the fermion mass. Indeed, for a fixed value of $n_{\text{eff}}$, a choice of the renormalisation scale $\mu_R = Q_0 / 2$ would exclude some values of the fermion mass $m_X \in (Q_0/2, 2Q_0)$ that would obviously not be excluded for $\mu_R = 2Q_0$. Thus, for this particular study, it seems doubtful that the factor of 2 prescription for the variations of $\mu_R, \mu_F$ is useful for estimating theoretical uncertainties.

\section{Conclusions}
\label{conclusions}

In this work, the dependence of $\alpha_s(Q^2)$ on the BSM model through TEEC, at approximate NNLO in pQCD has been studied. Specifically, BSM models that introduce a new coloured fermion to the SM, characterised through its mass, $m_X$, and the SU(3) representation under which it transforms, parametrised by the $n_{\text{eff}}$ parameter, have been analysed. The theoretical predictions relative to the SM has been studied for models with $n_{\text{eff}} = $ 1, 3, 5, and 15, and $m_X = $ 600 GeV as an example, highlighting the fact that such models can become very similar to the SM at low energies, necessitating study at high energies to truly distinguish between them. The comparison of the theory predictions with the experimental data collected by ATLAS at 13 TeV has been performed to obtain the value of $\text{CL}_\text{s}$ for such BSM models.\\
\newline
The agreement between experiment and theory is better for models where the additional fermion transforms under an SU(3) representation with lower $n_{\text{eff}}$ (for example, the SM with an additional quark, $n_{\text{eff}} = 1$) and higher $m_X$, some cases being undistinguishable from the SM at the energies available to date. Limits have been set for these BSM models at a 95\% confidence level, considering all points in the measurement phase space through a global fit, and restricting the study to high-energy events, $H_{T2} >$ 3500 GeV. Similar results are obtained for both datasets, with the former showing a slightly better performance for low values of $n_\text{eff}$.\\
\newline
Increases in the centre-of-mass energy from future colliders, as well as the improved precision of the data and eventual theoretical predictions at higher perturbative orders, would allow for the improvement of the analysis presented here by reducing errors in both data and theory, thus allowing to test new physics models for increasingly high masses.
\paragraph{\textbf{Acknowledgements}}
The authors would like to thank Rene Poncelet for useful discussions. This work was partially funded by the Community of Madrid (Spain) under the project 2022-T1/IND-23859.
\bibliographystyle{elsarticle-num-names} 
\bibliography{bibliography}



%\begin{thebibliography}{00}

%% For numbered reference style

%\bibitem{lamport94}
%  \textit{\LaTeX: a document preparation system},
%  Addison Wesley, Massachusetts,

%\end{thebibliography}
\end{document}